\documentclass[reprint, amsmath, amssymb, prb]{revtex4-2}

\usepackage{mathtools}
\usepackage{amssymb}
\usepackage{amsfonts}
\usepackage{amsmath}
\usepackage{graphicx}
\usepackage{dcolumn}
\usepackage{bm}
\usepackage{hyperref}
\usepackage{cases}
\usepackage{tikz-feynman}

\newcommand{\Article}{paper}

\newcommand{\MarkovChain}{Markov-Chain }
\newcommand{\MCMCLong}{\MarkovChain Monte Carlo }
\newcommand{\MCMC}{MCMC }
\newcommand{\MCMCpar}{(MCMC) }
\newcommand{\MCMCMCLong}{Many-Configuration \MCMCLong }
\newcommand{\MCMCMC}{MCMCMC }
\newcommand{\MCMCMCpar}{(MCMCMC hereafter)}

\newcommand{\Peq}{P_{\text{eq}}}
\newcommand{\Pprp}{P_{\text{prp}}}
\newcommand{\Pcond}[3]{{#1}({#2}\,|\,{#3})}
\newcommand{\Pacc}{P_{\text{acc}}}
\newcommand{\Pgen}{P_{\text{gen}}}
\newcommand{\Pnode}{P_{\text{node}}}
\newcommand{\Ptime}[2]{P(({#1},{#2}))}
\newcommand{\Psubset}{P_{\text{subset}}}

\newcommand{\DirectedGraph}{directed graph }
\newcommand{\UndirectedGraph}{undirected graph }

\begin{document}

\title{\MCMCMCLong}

\author{ Fedor \surname{\v{S}imkovic IV} } \email{fsimkovic@gmail.com}
\affiliation{CPHT, CNRS, Ecole Polytechnique, Institut Polytechnique
  de Paris, Route de Saclay, 91128 Palaiseau, France\\ Coll\`ege de
  France, 11 place Marcelin Berthelot, 75005 Paris, France}

\author{Riccardo \surname{Rossi}} \email{riccardorossi4@gmail.com}
\affiliation{ Center for Computational
  Quantum Physics, Flatiron Institute, 162 5th Avenue, New York, New
  York 10010, USA}

\affiliation{Institute of Physics, Ecole Polytechnique F\'ed\'erale de
  Lausanne (EPFL), CH-1015 Lausanne, Switzerland}


\date{\today}

\begin{abstract}
We propose a minimal generalization of the celebrated \MCMCLong algorithm which allows for an arbitrary number of configurations to be visited at every Monte Carlo step. This is advantageous when a parallel computing machine is available, or when many biased configurations can be evaluated at little additional computational cost. As an example of the former case, we report a significant reduction of the thermalization time for the paradigmatic Sherrington-Kirkpatrick spin-glass model. For the latter case, we show that, by leveraging on the exponential number of biased configurations automatically computed by Diagrammatic Monte Carlo, we can speed up computations in the Fermi-Hubbard model by two orders of magnitude.
\end{abstract}

\maketitle

\section{Introduction
  \label{sec_intro}}
The Markov-Chain Monte Carlo method~\cite{mcmc} \MCMCpar is a generic
algorithm that allows to draw samples from a given unnormalized
probability distribution. It has found applications in many areas of
science, in particular in classical and quantum physics~\cite{krauth_book,
pwerner_book, sorella_book}, and statistics~\cite{bayesian_mcmc_book}.

The Markov-chain thermalization time is one of the critical properties
of the MCMC algorithm~\cite{mixing_times_ref}. Also known as mixing time, it is
the time it takes for the Markov chain to approach the steady-state
distribution, which is sometimes called the thermal-equilibrium distribution in statistical
physics. In the context of a Monte Carlo calculation, the
thermalization time is the time one needs to wait before accumulating
statistics. If it is larger than the total number of Monte Carlo steps, the \MCMC technique is not applicable. Another critical property is the autocorrelation time, which is defined as the time it takes to draw two uncorrelated samples from the Markov chain, after
the thermalization time has passed.  It can be reduced linearly by
using a parallel computing machine running independent Markov chains,
while the thermalization time cannot be reduced indefinitely with this
approach.  Introducing couplings between Markov chains has been shown to help
reduce the thermalization time: one usually either considers chains at
different temperatures~\cite{replica_mc, mc3}, or constructs unbiased
estimators~\cite{coupling_mcmc}.  Another strategy to reduce
thermalization time is the use of irreversible Markov
chains~\cite{krauth_ecmcmc, manon_ecmcmc}, a method that has recently
been extended to allow for parallelization~\cite{krauth_parallel}.  On
the other hand, quasi-Monte Carlo techniques~\cite{quasi_mc_1} do not
have a thermalization time by construction, and have recently been
successfully applied to quantum impurity problems~\cite{quasi_qmc},
but their application to large configurations spaces has not yet been
fully explored.

In some situations, one is able to generate many biased proposals for
new configurations of the Markov chain with little or no additional
computational effort.  This is the case, for instance, with some variants
of the Diagrammatic Monte Carlo technique as they automatically generate an
exponential number of configurations at each Monte Carlo step at no
additional cost~\cite{cdet, fedor_sigma, alice_michel, rr_sigma,
Rossi2020, Simkovic2020b}. Here, the MCMC algorithm can only consider one
configuration at a given Monte Carlo step, and there is no way to use
the information of the multiple biased configurations that have been generated.

In this \Article, we propose a minimal generalization of the \MCMCLong
 algorithm, which we call Many-Configuration Markov-Chain Monte Carlo
 (MCMCMC), that allows to consider multiple configurations at the same
 Monte Carlo step. This guarantees a straightforward parallelization
 of the computation in order to reduce the thermalization time, as
 well as allowing for the use of many biased configurations that might be available
 at a given Monte Carlo step.

The document is organized as follows: In Sec.~\ref{sec_problem} we
formulate and motivate the problem which this work addresses; In
Sec.~\ref{sec_intuitive} we present an intuitive picture of the
\MCMCMC technique by considering the case of two and three configurations explicitely; In Sec.~\ref{sec_ising} we present benchmark results for the Sherrington-Kirkpatrick spin-glass
model, showing how the use of a parallel machine can drastically reduce the thermalization time; In Sec.~\ref{sec_ftc} we provide numerical results for the Fermi-Hubbard
model within Diagrammatic Monte Carlo using a specifically designed version of the \MCMCMC algorithm.

\section{Problem statement
\label{sec_problem}}
The \MCMC method is based on the generation of a sequence of states,
called Markov chain, such that the transition probability between
these states depends only on the current state. The goal of a \MCMC
algorithm is to build a Markov chain such that, for every initial
state, the probability of a given state converges to an arbitrary
``equilibrium'' probability distribution $\Peq$ in the long-time
limit. The Markov chain is usually built by proposing a new state
$c_1$ given the current state $c_0$ with a certain, arbitrary,
proposal probability distribution, $\Pcond{\Pprp}{c_1}{c_0}$, which is
easy to sample from; then, in the simplest form of \MCMC, the proposed
state is accepted or rejected according to the Metropolis rate
satisfying detailed balance: $\text{min}(1,
\Peq(c_1)\,\Pcond{\Pprp}{c_0}{c_1}/
(\Peq(c_0)\,\Pcond{\Pprp}{c_1}{c_0}))$.  While being an extremely
general and powerful algorithm, there are situations where the
sequential generation of new configurations leads to inefficiency.  In
the following, we will consider two such situations:
\begin{enumerate}
  \item {A parallel computing machine is available, and the
  thermalization time for the Markov chain is non-negligeable }
\item {It is possible to compute a large number of biased configurations with little or no additional computational effort}
\end{enumerate}
In this \Article, we present a solution to the following problem: {\it
What is a minimal modification of \MCMCLong that allows to take
advantage of one (or both) of these possibilities?  }

\begin{figure}[t]
  \centering \includegraphics[width=0.4\textwidth]{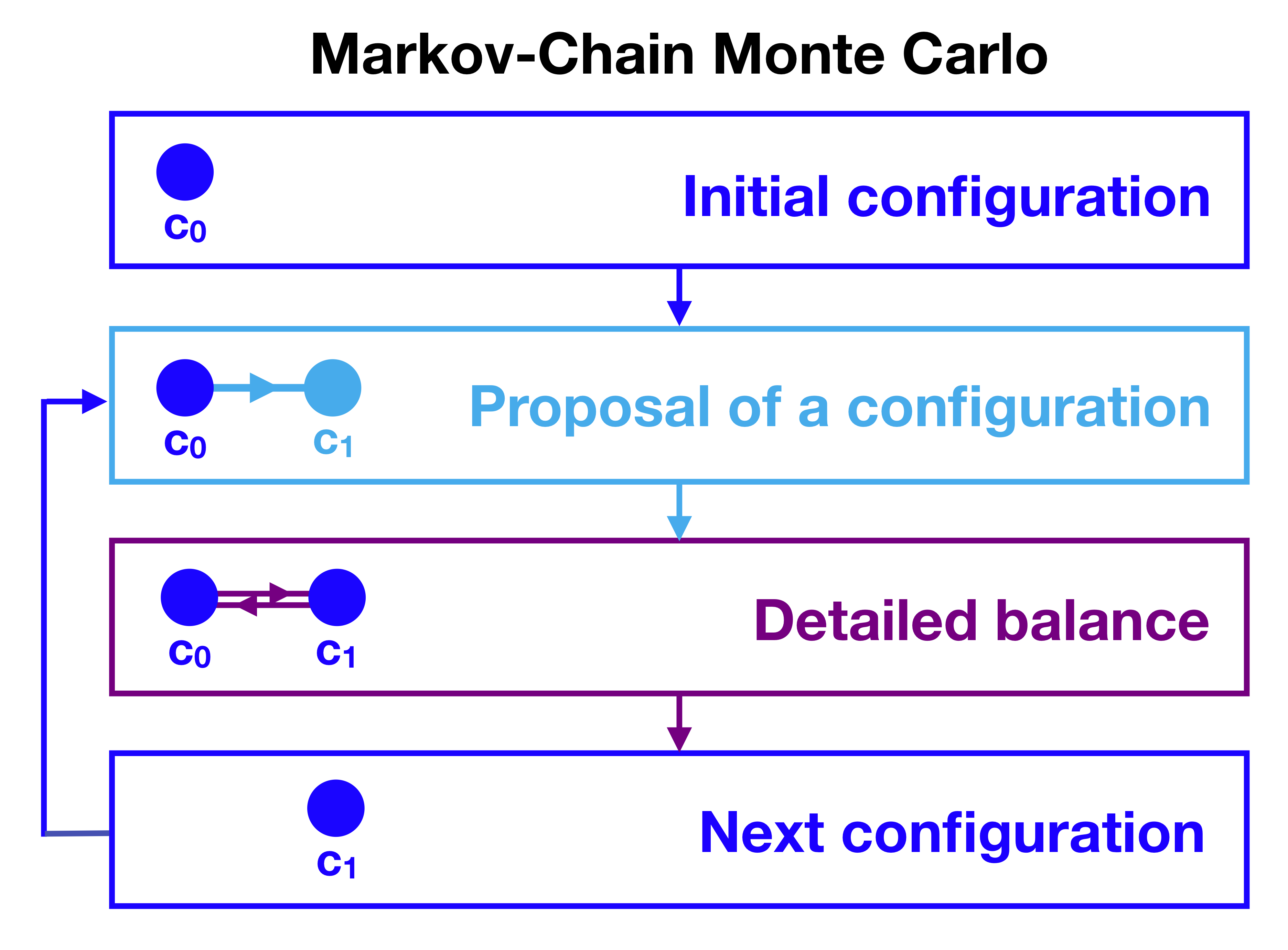} \caption{
  Construction of the Markov chain in standard MCMC.  The initial
  state of the Markov chain is $c_0$. The new proposed state is $c_1$;
  we depict this proposal process as a directed graph.  Then, detailed
  balance is imposed to stochastically select the next state of the
  Markov chain.  \label{fig_mcmc_intuitive}}
\end{figure}

As already mentioned, the difficulty of MCMC in dealing with these
situations can be traced back to its sequential definition: The Markov
chain is built by considering only {one} new configuration at a given
Monte Carlo step, and there is no way to simultaneously use the
information from multiple configurations.  Even if multiple
configurations can be generated, only at most one of them can be
accepted.  In particular, in the presence of non-negligeable
thermalization time, there is little advantage in using a parallel
computing machine over a single-core machine as any statistics
accumulated before thermalization has completed cannot be used.


\section{\label{sec_intuitive}
  Intuitive picture}
Below, we aim at providing the reader with an intuitive picture of the
novel algorithm presented in this work,
\MCMCMCLong \MCMCMCpar, in the graph version [the set version is discussed in the Appendix, see Sec.~\ref{sub_mcmc_set}]. Here, we do not provide complete derivations and proofs, all of which can be found in Appendix \ref{sec_definition}.

\begin{figure}
  \centering \includegraphics[width=0.4\textwidth]{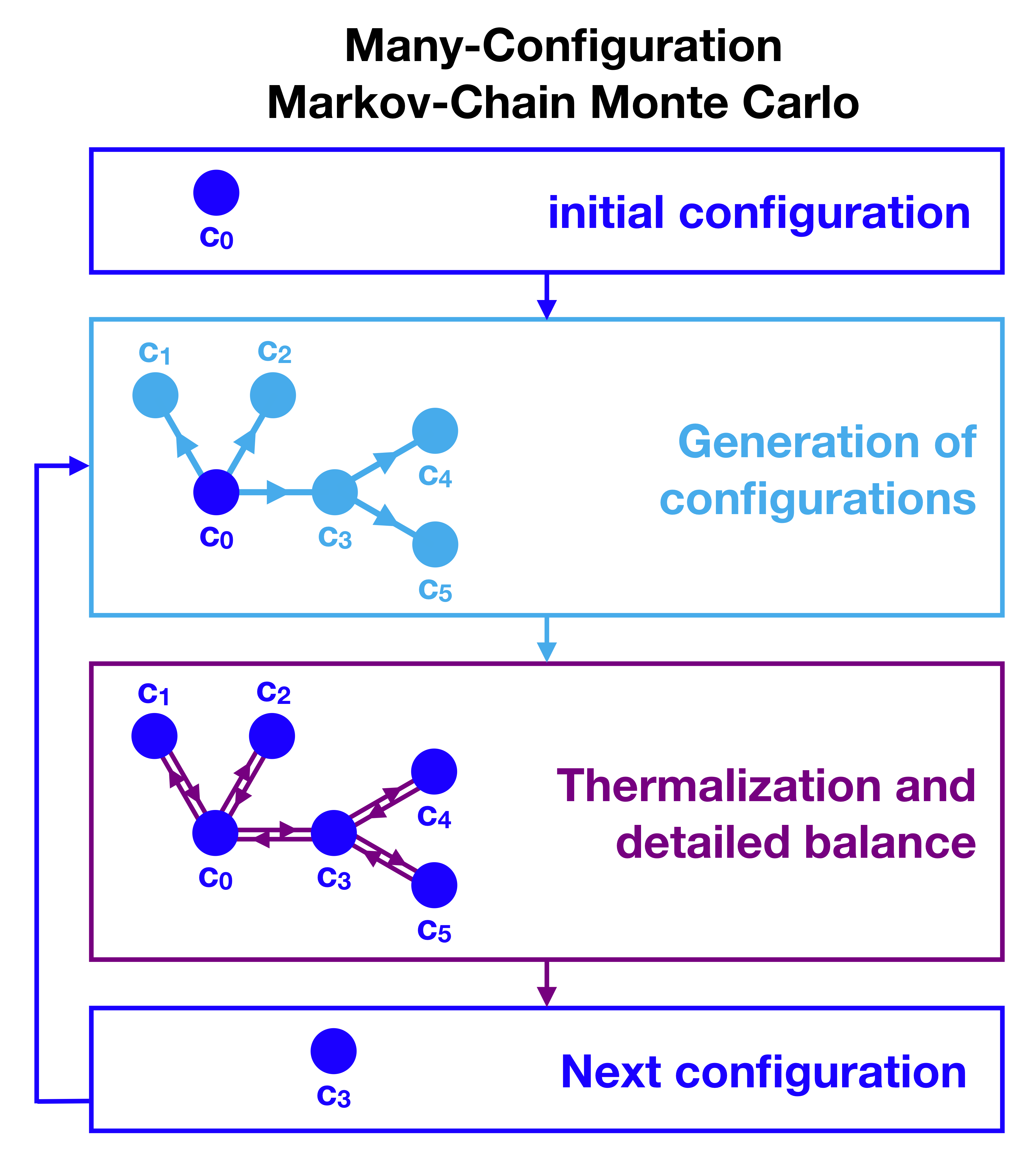} \caption{
  Construction of the Markov chain for MCMCMC.  The initial state of
  the Markov chain is $c_0$. The new generated configurations are
  $c_1, \dots, c_5$; we depict this generation process as a directed
  graph.  In the thermalization phase, the double-arrow graph is
  considered, and a fraction of a time step that allows to respect
  detailed balance within the graph is spent in each node of the
  graph. Then, the new state of the Markov chain is stochastically
  selected within the graph with a rate equal to the fraction of time
  spent in each node.  \label{fig_mcmcmc_intuitive}}
\end{figure}

Let us first describe a standard version of \MCMC in a language that
will easily allow for its generalization. Using the notation of
Fig.~\ref{fig_mcmc_intuitive}, let us suppose that, at a given Monte
Carlo step, the Markov-chain state is the configuration $c_0$, which
belongs to some configuration space we want to sample in. We propose a
new configuration $c_1$ as the next state of the Markov chain, where
$c_1$ is proposed given $c_0$ with some probability distribution
$\Pcond{\Pprp}{c_1}{c_0}$. $\Pprp$ can be freely chosen as long as it
is guaranteed that the Markov chain is ergodic. We schematically
depict this proposal process with a \DirectedGraph composed of the
nodes $c_0$ and $c_1$, and of an arrow connecting $c_0$ to $c_1$ [see
Fig.~\ref{fig_mcmc_intuitive}]. In order for the Markov chain to
converge to a given equilibrium distribution $\Peq$, it is sufficient
to impose the detailed-balance condition:
\begin{equation}\label{eq_detbal_metro_intuitive}
  \begin{split}
  &\Peq(c_0)\,\Pcond{\Pprp}{c_1}{c_0}\,\Pcond{\Pacc}{c_1}{c_0}=\\
  &\Peq(c_1)\,\Pcond{\Pprp}{c_0}{c_1}\,\Pcond{\Pacc}{c_0}{c_1},
  \end{split}
\end{equation}
where $\Pcond{\Pacc}{c_j}{c_k}$ is the probability of accepting the
proposed configuration $c_k$ as the next Markov-chain state when
proposing it from the configuration $c_j$. In
Fig.~\ref{fig_mcmc_intuitive} we represent the detailed-balance
process by an undirected graph with double arrows. After the
acceptance/rejection process, the Markov-chain state is either $c_0$
or $c_1$, and this whole procedure is repeated at the next Monte Carlo
step.

We now introduce the \MCMCMC algorithm as a natural many-configuration
generalization of MCMC. Let $c_0$ be the Markov-chain state at a given
Monte Carlo time step. We want to follow as closely as possible
the \MCMC procedure illustrated in Fig.~\ref{fig_mcmc_intuitive}. For
this reason, we want to find a way to propose multiple configurations
by repeatedly using the ``two-body'' probability distribution
$\Pcond{\Pprp}{c_k}{c_j}$ introduced above. A natural way to achieve
this is sketched in Fig.~\ref{fig_mcmcmc_intuitive}: We consider the
configurations $c_0,\dots,c_5$ to be nodes of a directed graph; an
arrow going from $c_j$ to $c_k$ means that the node $c_k$ has been
proposed using the probability distribution
$\Pcond{\Pprp}{c_k}{c_j}$. The graph structure and generation process are completely
arbitrary; in the following we will detail some of the possible
choices. In the \MCMC procedure described above, once the graph is
generated, one of the two configurations is chosen as the Markov-chain
state at the next step. In the case of \MCMCMC, as the graph we consider
at each step can be very big, we want to make full use of all the
proposed configurations. For this reason, we add a thermalization
phase: All the nodes of the graph are visited a fraction of time
proportional to the rate at which they would be chosen according to
the detailed-balance condition between configurations belonging to the
same undirected graph [see Fig.~\ref{fig_mcmcmc_intuitive}]. After
this thermalization phase, a node of the graph is chosen with a
probability proportional to the rate in which it is visited, and the
process is iterated.

In the rest of this section we want to progressively motivate \MCMCMC
as a minimal extension of \MCMC by considering the case of three
proposed configurations per Monte Carlo step in detail.  For this
reason, we are going to present three algorithms that ``interpolate''
between \MCMC and \MCMCMC: the first is a rewriting of the standard
\MCMC with a heat-bath acceptance rate within the graph language we
use in this \Article, the second is a modification of the first where
we visit multiple configurations at each Monte Carlo step, and the
third is a \MCMCMC algorithm for many configurations that
we specialize to the case of three configurations.

\subsection{\label{sub_intuitive_mcmc_graph}\MCMC in the language of graphs}
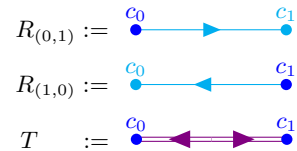
\begin{figure}
  \centering
  $R_{(0,1)} :=$
  \color{cyan}
  \begin{tikzpicture}
    \begin{feynman}
      \color{blue}
      \vertex [dot, label=$c_0$] (a) at (0,0);
      \color{cyan}
      \vertex [dot, label=$c_1$] (b) at (2,0);]
        \color{cyan}
    \node at (b) [circle,fill,inner sep=1.5pt]{};
    \diagram*{
    (a) -- [fermion] (b)
    };
    \color{blue}
        \node at (a) [circle,fill,inner sep=1.5pt]{};
    \end{feynman}
  \end{tikzpicture}

  \vspace{1mm}
  \color{black}
  $R_{(1,0)} :=$
  \begin{tikzpicture}
    \begin{feynman}
      \color{cyan}
      \vertex [dot, label=$c_0$] (a) at (0,0);
            \color{blue}
            \vertex [dot, label=$c_1$] (b) at (2,0);]
                    \color{cyan}
    \node at (a) [circle,fill,inner sep=1.5pt]{};
    \diagram*{
    (a) -- [anti fermion] (b)
    };
    \color{blue}
        \node at (b) [circle,fill,inner sep=1.5pt]{};
    \end{feynman}
  \end{tikzpicture}

  \vspace{1mm}
    \color{black}
    $\;\;T_{\phantom{(0,0)}}:=$
\begin{tikzpicture}
  \begin{feynman}
    \color{blue}
      \vertex [dot, label=$c_0$] (a) at (0,0);
      \vertex [] (c) at (1,0);]
        \vertex [dot, label=$c_1$] (b) at (2,0);]
          \color{violet}
    \diagram*{
    (c) --[double,double distance=0.3ex,with arrow=0.4,arrow size=0.2em]   (a)
    };
    \diagram*{
    (c) --[double,double distance=0.3ex,with arrow=0.4,arrow size=0.2em]   (b)
    };
    \color{blue}
        \node at (a) [circle, fill, inner sep=1.5pt]{};
    \node at (b) [circle, fill, inner sep=1.5pt]{};
    \end{feynman}
  \end{tikzpicture}
\color{black}
  \caption{Top: in $R_{(0,1)}$ the configuration $c_1$ is generated from $c_0$.
    Center: in $R_{(1,0)}$ the configuration $c_0$ is generated from $c_1$.
    Bottom: the undirected graph $T$ describes the two processes.
  \label{fig_two_conf}}
\end{figure}

We now discuss in more detail the \MCMC algorithm of
Fig.~\ref{fig_mcmc_intuitive}, highlighting its graphical
interpretation. We suppose that, at a given Monte Carlo step, the
Markov-chain state is the configuration $c_0$. As discussed above, we
propose the configuration $c_1$ as the next Markov-chain state given
$c_0$ with the probability distribution $\Pcond{\Pprp}{c_1}{c_0}$. We
can represent this process by a \DirectedGraph $R_{(0,1)}$ with two
nodes, $c_0$ and $c_1$, and an arrow connecting the configuration $c_0$
to the configuration $c_1$, as illustrated in Fig.~\ref{fig_two_conf}.

As we want to use detailed balance, we also need to consider the
inverse process in which the configuration $c_0$ is proposed from
$c_1$. We can represent the inverse process by a \DirectedGraph
$R_{(1, 0)}$ with the same nodes as $R_{(0, 1)}$ and opposite
direction for the arrow [see Fig.~\ref{fig_two_conf}].
Finally, we introduce the \UndirectedGraph $T$ having the same nodes as
$R_{(0, 1)}$ and $R_{(1, 0)}$ [see Fig.~\ref{fig_two_conf}].  We
define $\Pcond{\Pacc}{c}{T}$ as the probability of accepting a node $c
\in \{c_0,c_1\}$ of the undirected graph $T$ as the next state of the
Markov chain, which, for simplicity, we choose to be independent on
the creation history of $T$.  We decide to impose the detailed balance
condition independently on any such $T$:
\begin{equation}\label{eq_detbal_two_conf}
\begin{split}
  &\Peq(c_0)\,
  \Pcond{\Pprp}{T}{c_0}\,
  \Pcond{\Pacc}{c_1}{T}=\\
  &\Peq(c_1)\,
  \Pcond{\Pprp}{T}{c_1}\,
  \Pcond{\Pacc}{c_0}{T},\\
\end{split}
\end{equation}
where we defined $\Pcond{\Pprp}{T}{c_\alpha}
:= \Pcond{\Pprp}{c_{1-\alpha}}{c_\alpha}$ as the probability of
proposing $c_{1-\alpha}$ given $c_\alpha$, as discussed above.  We
remark that while Eq.~\ref{eq_detbal_two_conf} is equivalent to
Eq.~\ref{eq_detbal_metro_intuitive}, their interpretation is different: In
Eq.~\ref{eq_detbal_metro_intuitive} we have directly considered the
detailed balance between configurations, while in
Eq.~\ref{eq_detbal_two_conf} we have used the graph $T$ as
an intermediary.  One can easily show that the following choice of
$\Pacc$ satisfies the detailed balance condition
[Eq.~\eqref{eq_detbal_two_conf}]:
\begin{equation}\label{eq_node_two_conf}
  \Pcond{\Pacc}{c}{T}:=
\frac{\Peq(c)\,\Pcond{\Pprp}{T}{c}}
{\sum_{c^\prime\in V(T)}\Peq(c^\prime)\,\Pcond{\Pprp}{T}{c'}},
\end{equation}
where $V(T):=\{c_0,c_1\}$ is the set of nodes of $T$.  This is
equivalent to standard MCMC with a ``heat-bath'' acceptance rate.

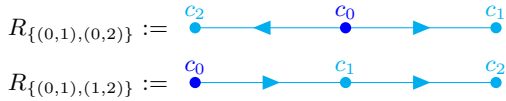
\begin{figure}[t]
  \centering
  $R_{\{(0,1),(0,2)\}}:=$
    \color{cyan}
  \begin{tikzpicture}
    \begin{feynman}
      \vertex [dot, label=$c_2$, dot] (a) at (0,0);
      \color{blue}
      \vertex [dot, label=$c_0$] (b) at (2,0);
      \color{cyan}
    \vertex [dot, label=$c_1$] (c) at (4,0);]
    \node at (a) [circle,fill,inner sep=1.5pt]{};
    \node at (c) [circle,fill,inner sep=1.5pt]{};
    \diagram*{
    (a) -- [anti fermion] (b) -- [fermion] (c)
    };
    \color{blue}
        \node at (b) [circle,fill,inner sep=1.5pt]{};
    \end{feynman}
  \end{tikzpicture}

  \vspace{1mm}
  \color{black}
$R_{\{(0,1),(1,2)\}}:=$
  \begin{tikzpicture}
    \begin{feynman}
      \color{blue}
      \vertex [dot, label=$c_0$, dot] (a) at (0,0);
      \color{cyan}
      \vertex [dot, label=$c_1$] (b) at (2,0);
    \vertex [dot, label=$c_2$] (c) at (4,0);]
    \node at (b) [circle,fill,inner sep=1.5pt]{};
    \node at (c) [circle,fill,inner sep=1.5pt]{};
    \diagram*{
    (a) -- [fermion] (b) -- [fermion] (c)
    };
    \color{blue}
    \node at (a) [circle,fill,inner sep=1.5pt]{};
    \end{feynman}
  \end{tikzpicture}

  \color{black} \caption{The two directed graphs of configurations
  that can be generated starting from $c_0$.  \label{fig_three_conf}}
\end{figure}

\subsection{Generalization to three configurations per Monte Carlo step}
In this section we generalize the procedure of
Sec.~\ref{sub_intuitive_mcmc_graph} to the case of three
configurations per Monte Carlo step, for a particular graph-generation
process.  The main purpose of this section is to give the intuition
behind the procedure, without providing a proof for all the
statements, which is left to the Appendix [see
Sec.~\ref{sub_mcmc_graph}].

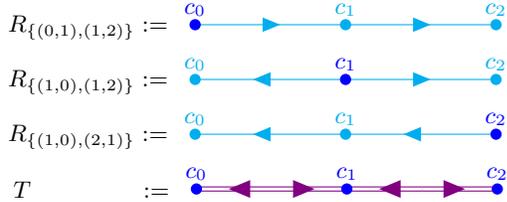
\begin{figure}
  \centering
  \vspace{1mm}
$R_{\{(0,1),(1,2)\}}:=$
  \begin{tikzpicture}
    \begin{feynman}
      \color{blue}
      \vertex [dot, label=$c_0$, dot] (a) at (0,0);
        \color{cyan}
    \vertex [dot, label=$c_1$] (b) at (2,0);
    \vertex [dot, label=$c_2$] (c) at (4,0);]
    \node at (b) [circle,fill,inner sep=1.5pt]{};
    \node at (c) [circle,fill,inner sep=1.5pt]{};
    \diagram*{
    (a) -- [fermion] (b) -- [fermion] (c)
    };
    \color{blue}
        \node at (a) [circle,fill,inner sep=1.5pt]{};
    \end{feynman}
  \end{tikzpicture}

  \vspace{1mm}
    $R_{\{(1,0),(1,2)\}}:=$
  \begin{tikzpicture}
    \begin{feynman}
      \color{blue}
      \vertex [dot, label=$c_1$] (b) at (2,0);
            \color{cyan}
      \vertex [dot, label=$c_0$, dot] (a) at (0,0);
    \vertex [dot, label=$c_2$] (c) at (4,0);]
      \node at (a) [circle,fill,inner sep=1.5pt]{};
    \node at (c) [circle,fill,inner sep=1.5pt]{};
    \diagram*{
    (a) -- [anti fermion] (b) -- [fermion] (c)
    };
    \end{feynman}
    \color{blue}
        \node at (b) [circle,fill,inner sep=1.5pt]{};
  \end{tikzpicture}

  \vspace{1mm}

$R_{\{(1,0),(2,1)\}}:=$
  \begin{tikzpicture}
    \begin{feynman}
      \color{blue}
      \vertex [dot, label=$c_2$] (c) at (4,0);]
      \color{cyan}
    \vertex [dot, label=$c_0$, dot] (a) at (0,0);
    \vertex [dot, label=$c_1$] (b) at (2,0);
    \node at (a) [circle,fill,inner sep=1.5pt]{};
    \node at (b) [circle,fill,inner sep=1.5pt]{};
    \diagram*{
    (a) -- [anti fermion] (b) -- [anti fermion] (c)
    };
    \end{feynman}
    \color{blue}
        \node at (c) [circle,fill,inner sep=1.5pt]{};
      \end{tikzpicture}

  \vspace{1mm}
  $\;T_{\phantom{\{(0,1),(0,2)\}}}:=$
  \color{violet}
  \begin{tikzpicture}
    \begin{feynman}
      \color{blue}
      \vertex [dot, label=$c_0$, dot] (a) at (0,0);
          \vertex [] (d) at (1,0);
          \vertex [dot, label=$c_1$] (b) at (2,0);
          \vertex [] (e) at (3,0);
          \vertex [dot, label=$c_2$] (c) at (4,0);]
            \color{violet}
        \diagram*{
    (d) --[double,double distance=0.3ex,with arrow=0.4,arrow size=0.2em]   (a)
        };
        \diagram*{
    (d) --[double,double distance=0.3ex,with arrow=0.4,arrow size=0.2em]   (b)
        };
        \diagram*{
    (e) --[double,double distance=0.3ex,with arrow=0.4,arrow size=0.2em]   (b)
        };
        \diagram*{
    (e) --[double,double distance=0.3ex,with arrow=0.4,arrow size=0.2em]   (c)
        };
        \color{blue}
        \node at (a) [circle,fill,inner sep=1.5pt]{};
    \node at (b) [circle,fill,inner sep=1.5pt]{};
    \node at (c) [circle,fill,inner sep=1.5pt]{};
    \end{feynman}
  \end{tikzpicture}

  \color{black}
  \caption{Directed graphs that correspond to the same undirected graph $T$.
  \label{fig_rooted_trees_three}}
\end{figure}

Suppose that, at a given Monte Carlo step, the Markov-chain state is
$c_0$.  We propose a new configuration $c_1$ with probability
$\Pcond{\Pprp}{c_1}{c_0}$, analogously to what is done in
Sec.~\ref{sub_intuitive_mcmc_graph}. At this point, there are two
possibilities: With probability $p$, we generate a new configuration
$c_2$ from $c_0$ with generation probability
$\Pcond{\Pprp}{c_2}{c_0}$; otherwise, with probability $q=1-p$, we
generate a new configuration $c_2$ from $c_1$ with generation
probability $\Pcond{\Pprp}{c_2}{c_1}$. We represent these two
processes by two directed graphs, which we respectively denote by
$R_{\{(0,1),(0, 2)\}}$ and $R_{\{(0,1),(1,2)\}}$ [see
Fig.~\ref{fig_three_conf}]. Suppose that we have stochastically chosen
to generate the first directed graph in Fig.~\ref{fig_three_conf},
$R_{\{(0,1),(1,2)\}}$. Let $T$ be the undirected graph with the same
nodes as $R_{\{(0,1),(1,2)\}}$ [see
Fig.~\ref{fig_rooted_trees_three}].  We consider all the directed
graphs that have the same nodes as $T$, and edges that can be
obtained from $T$ by removing one arrow per edge: these are
$R_{\{(0,1),(1,2)\}}$, $R_{\{(1,0),(1,2)\}}$, and
$R_{\{(1,0),(2,1)\}}$; they correspond to the different ways of
generating the graph $T$ starting from one of the nodes [see
Fig.~\ref{fig_rooted_trees_three} for the definition].

As the next Markov-chain state, we choose $c \in \{c_0,c_1,c_2\}$ with
probability $\Pcond{\Pacc}{c}{T}$, a quantity that only depends on
the undirected graph $T$. In order to have a useful Monte Carlo
process, we need to impose that the average time spent in a given
configuration $c$ coincides with $\Peq(c)$. For this, it is sufficient
to impose the detailed balance condition:
\begin{equation}\label{eq_detbal_two_conf}
\begin{split}
  &\Peq(c_j)\,
  \Pcond{\Pprp}{T}{c_j}\,
  \Pcond{\Pacc}{c_k}{T}=\\
  &\Peq(c_k)\,
  \Pcond{\Pprp}{T}{c_k}\,
  \Pcond{\Pacc}{c_j}{T},\\
\end{split}
\end{equation}
where $c_j$ and $c_k$ are nodes of the undirected graph $T$.

It can be shown [see Sec.~\ref{sub_mcmc_graph}] that the following
choice satisfies detailed balance:
\begin{equation}\label{eq_detbal_intuitive}
\Pcond{\Pacc}{c}{T}
:=
\frac{
\Peq(c)\,
\Pcond{\Pprp}{T}{c}
}
{
\sum_{c'\in V(T)}
\Peq(c')\,
\Pcond{\Pprp}{T}{c'}
},
\end{equation}
where $V(T):=\{c_0,c_1,c_2\}$ is the set of nodes of the graph
$T$, and $\Pcond{\Pprp}{T}{c}$ is the probability of generating
the graph $T$ from $c$, which, for the specific stochastic graph
generation process we consider in this section, is given by
\begin{equation}\label{}
\begin{split}
&\Pcond{\Pprp}{T}{c_0}=
\Pcond{\Pprp}{c_1}{c_0}\,
q\,
\Pcond{\Pprp}{c_2}{c_1}\\
&\Pcond{\Pprp}{T}{c_1}=
2\,\Pcond{\Pprp}{c_0}{c_1}\,
p\,
\Pcond{\Pprp}{c_2}{c_1}\\
&\Pcond{\Pprp}{T}{c_2}=
\Pcond{\Pprp}{c_1}{c_2}\,
q\,
\Pcond{\Pprp}{c_0}{c_1},\\
\end{split}
\end{equation}
where we have taken into account the two ways of generating $T$ given
$c_1$.

\subsection{\label{sub_intuitive_cont}Visiting many configurations in one Monte Carlo step by ``thermalization'' of the undirected graph}
\begin{figure}
  \centering \includegraphics[width=0.47\textwidth]{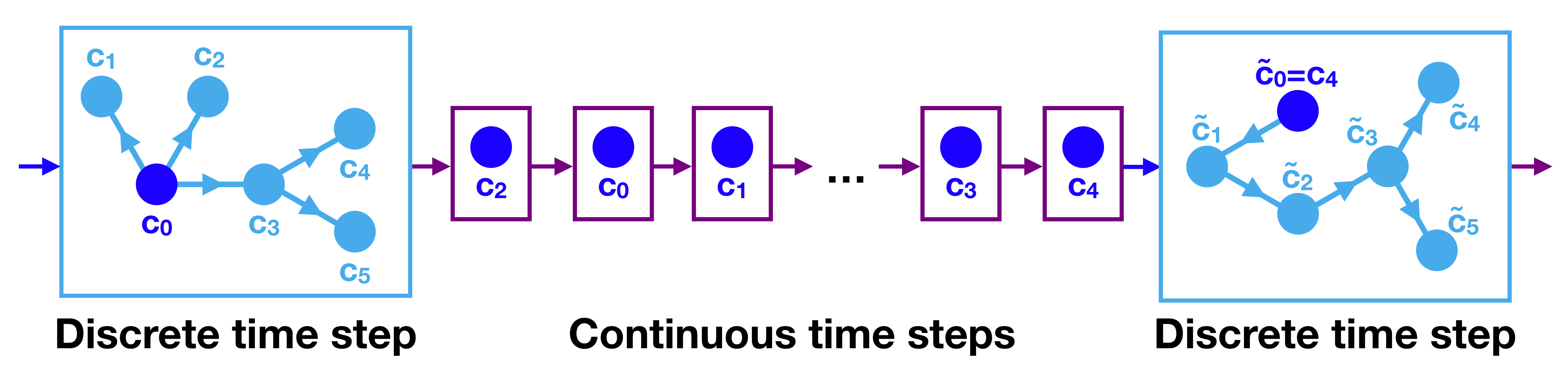} \caption{
 Discrete and continuous Monte Carlo time steps in MCMCMC. A discrete
 step corresponds to the generation of a new graph of configurations.
 A continuous step is a virtual update between two configurations
 which are connected by an edge within a given graph, a process which
 we refer to as ``thermalization'' of the undirected
 graph. \label{fig_time_steps}}
\end{figure}

In order to render the algorithm more efficient when many
configurations are simultaneously available, we introduce a
modification of the Markov chain definition that allows to visit
multiple configurations per Monte Carlo step. We call this extension
the ``thermalization'' of the undirected graph [see
Fig.~\ref{fig_time_steps}]. As an intuitive picture, we can imagine that
the generated undirected graph of configurations is the whole
configuration space for one time step, and continuous-time Monte Carlo
updates are made between nodes of the graph. Formally, this will be
simply done by renaming the ``probability of proposal of
configurations'', $\Pprp$, to the ``probability of generation of
configurations'', $\Pgen$, and interpreting the probability of
acceptance of a configuration $\Pacc$ as the fraction of time spent in
a node of the graph $\Pnode$.

A powerful way to formalize this concept is by considering a Markov
chain of undirected graphs instead of the usual Markov chain of
configurations. Suppose that at a given Monte Carlo step the
Markov-chain state is the undirected graph $T$ consisting of the nodes
$c_0$ and $c_1$ [see Fig.~\ref{fig_two_conf}]. We generate the next
state of the Markov chain, $T'$, in the following way: We choose one
of the nodes of $T$, say $c_0$, with a certain probability
distribution $\Pcond{\Pnode}{c_0}{T}$ that coincides with
$\Pcond{\Pacc}{c_0}{T}$ [defined in
Sec.~\ref{sub_intuitive_mcmc_graph}]; We then generate the new state
of the Markov chain, $T'$, with a probability $\Pcond{\Pgen}{T'}{c_0}$
that coincides with $\Pcond{\Pprp}{T'}{c_0}$ [$\Pprp$ is introduced in
Sec.~\ref{sub_intuitive_mcmc_graph}]. To this discrete-time Markov
chain of undirected graphs, we associate a Markov chain of
configurations in continuous time in the following way: Suppose that
the Markov-chain state at discrete time $j$ is $T$; For a continuous
time $t$ such that $j \le t < j+1$, we spend a fraction of the time equal to
$\Pcond{\Pnode}{c}{T}$ in a given configuration $c$ of the undirected graph $T$ [see Fig.~\ref{fig_time_steps} for an intuitive
picture]. For any given $\Pgen\equiv \Pprp$, we adjust $\Pnode$ to
achieve an average time of $\Peq(c)$ spent in each configuration
$c$. It can be shown that substituing $\Pacc$ with $\Pnode$ and
$\Pprp$ with $\Pgen$ in Eq.~\eqref{eq_node_two_conf} guarantees this
equilibrium condition. In Sec.~\ref{sub_mcmc_graph} we provide a
formal proof of this fact.

\section{\label{sec_ising}
  Application to a spin-glass model}
\subsection{\label{sub_ising}The Sherrington-Kirkpatrick model}
We now present the implementation of the \MCMCMC algorithm for the
Sherrington-Kirkpatrick model, defined by the energy
\begin{equation}\label{}
  E(\{J_{jk}\},\{S_j\}):=
  \frac{1}{\sqrt{L}}
  \sum_{0\le j<k\le L-1}
  J_{jk}\;S_j\,S_k
\end{equation}
where $L$ is the system size, $S_j\in\{-1,1\}$, and
$J_{jk}\in\mathbb{R}$ are random gaussian numbers of zero mean and
unit variance. We consider the average energy per site:
\begin{equation}\label{}
  \frac{\langle E\rangle}{L}:=
  \frac{  \sum_{ \{S_j\} }
    e^{-E(\{J_{jk}\},\{S_j\})/T}\;
    E(\{J_{jk}\},\{S_j\})/L  }
       {  \sum_{ \{S_j\} }
         e^{-E(\{J_{jk}\},\{S_j\})/T}  }
\end{equation}
where $T$ is the temperature.
The model can be solved exactly in the thermodynamic
limit~\cite{parisi_sk}, and at the critical temperature $T_c=1$ the
system undergoes a replica-symmetry-breaking phase transition. The
motivation to study it numerically is the fact that close but above
the critical temperature, the system thermalizes very slowly, and can
therefore be used as a benchmark for our technique. We consider here
directly the critical case $T=T_c$.


\subsection{\label{sub_ising_results}Numerical results}
We have compared the following algorithms for the
Sherrington-Kirkpatrick model simulation: the simplest local MCMC
algorithm; the \MCMCMC extension of the local
\MCMC with the line-graph addition process [see Sec.~\ref{sub_line_graph}
  for details]; and the graph populating process for a $k$-ary tree
 [see Sec.~\ref{sub_gen_graph} for details].

\begin{figure}
  \centering \includegraphics[width=0.5\textwidth]{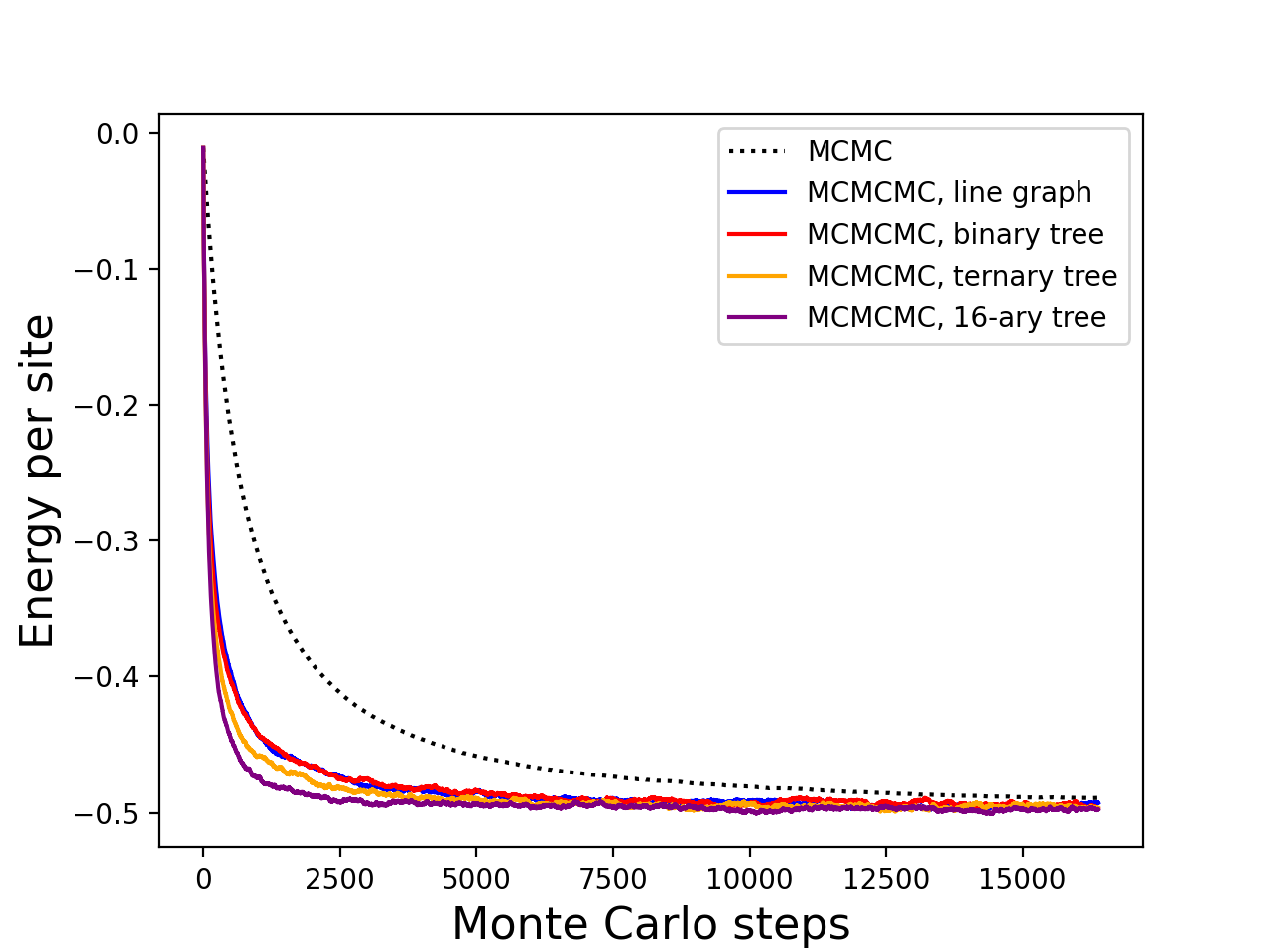} \caption{Energy
  per site, averaged over parallel computing cores, for a particular
  disorder realization of the Sherrington-Kirkpatrick model with
  $L=2^{10}$ and $T=1$, as a function of Monte Carlo steps. The \MCMC
  dotted line represents the standard local Metropolis algorithm,
  while the solid lines are the MCMCMC results. The blue line is the
  line-graph random addition process with graph size of $2^{7}$, the
  red line is the binary tree of depth $7$, the yellow line is the
  ternary tree of depth $5$, the violet line is the 16-ary tree of
  depth $3$.
  \label{fig_sk}}
\end{figure}

In Fig.~\ref{fig_sk}, we present the results of the numerical
comparison between the standard local Metropolis \MCMC algorithm with
a single update, and various \MCMCMC techniques differing by graph
structure.  We see that the thermalization time is decreased by two
orders of magnitude, proving therefore the success of the algorithm in
dramatically reducing the thermalization time.

\section{\label{sec_ftc}Application to lattice fermions}
\subsection{\label{sub_diagmc}The Fermi-Hubbard model}
We consider the doped two-dimensional fermionic Hubbard
model \cite{hubbard1963electron, anderson1963theory,
anderson1997theory}:
\begin{equation}
H = \sum_{k,\sigma} \left(\epsilon_{k} -\mu\right)c_{k\sigma}^\dagger
c_{k\sigma}+U\sum_i n_{i\uparrow}n_{i\downarrow}.
\label{H}
\end{equation}
Here $\mu$ denotes the chemical potential, $k$ the momentum,
$\sigma \in \{ \uparrow, \downarrow \} $ the spin, $U$ the onsite
repulsion strength, $i$ labels lattice sites, and the (square lattice)
dispersion is given by
\begin{align}
\epsilon_k=-2t\left[\cos(k_x)+\cos(k_y)\right],
\end{align}
where $t$ is the nearest-neighbor hopping amplitude (we set $t=1$ in
our units). We further define $T$ as the temperature and $n$ as the
density.

In the chosen parameter regime of $T=0.1$ and $n=0.875$, one needs
around ten well-computed expansion coefficients for the double
occupancy (errorbars below $1\%$) to be able to resum the resulting
series at interaction strength $U=8$, which represents a much
investigated strongly correlated regime of the model with stripe
magnetic order in the ground state \cite{leblanc2015solutions,
Zheng1155, qin2020absence}.

\subsection{\label{sub_diagmc_results}Numerical results}
As traditional Quantum Monte Carlo techniques are affected from the
fermionic sign problem when simulating the repulsive Hubbard model
away from half filling, we use Diagrammatic Monte Carlo, which allows
to circumvent this issue by sampling Feynman diagrams directly in the
thermodynamic limit.  We hereafter use a \MCMCMC algorithm for sets we
specifically developed for use within the Connected Determinant
Diagrammatic Monte Carlo (CDet)~\cite{cdet} [see
Sec.~\ref{sec_intro_diagmc} for a CDet introduction].  The reader can
find more details about the set generating process we use in the
Appendix, Sec.~\ref{sec_generation_sets}.

\begin{figure}[hb]
  \centering
  \includegraphics[width=0.4\textwidth]{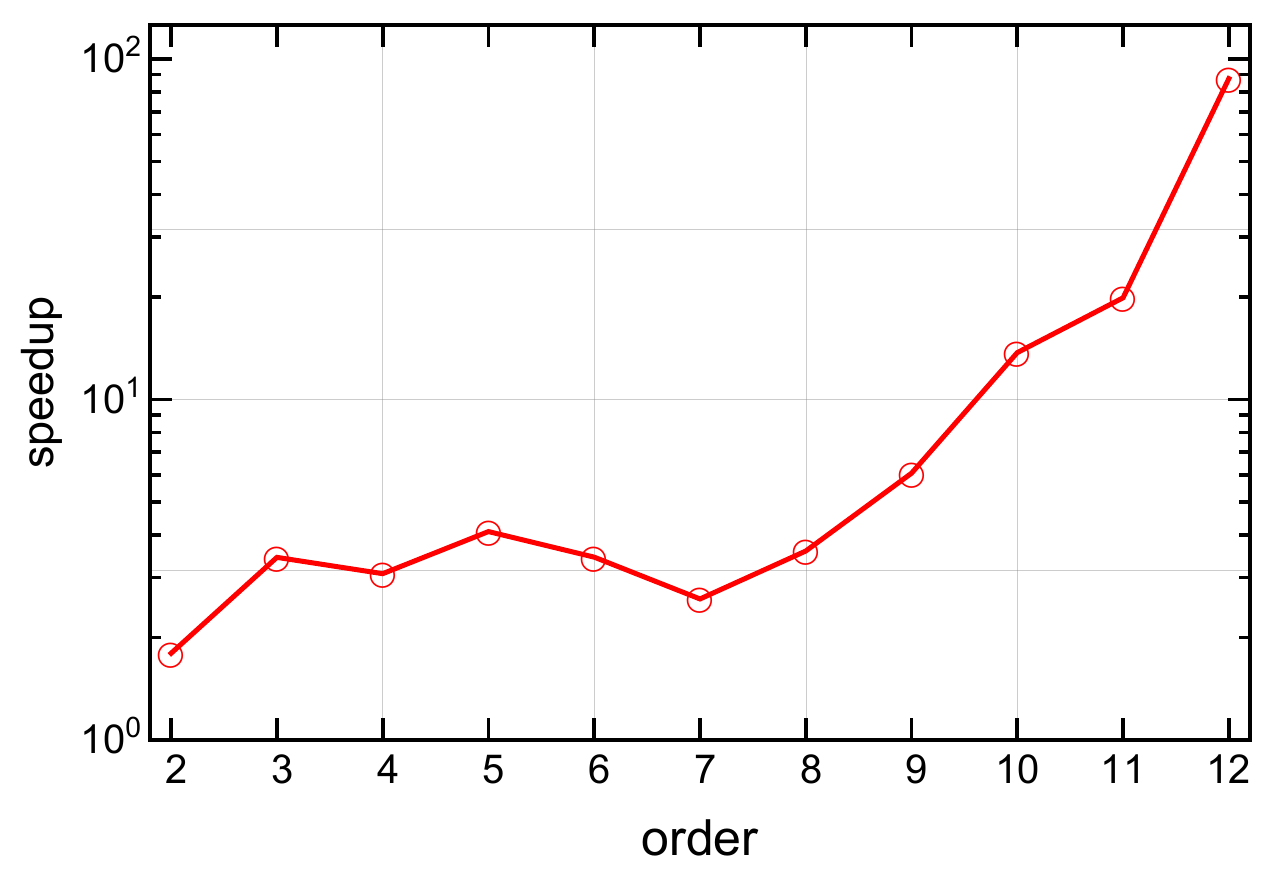}

  \caption{Improvement in computational time needed to reach a given
    stochastic error for a given expansion coefficient of the double
    occupancy as a function of expansion order for the Fermi-Hubbard
    model in the thermodynamic limit, at $T=0.1$ and $1/8$ doping.
  \label{fig_cdet_benchmark}}
\end{figure}

To assess relative performance of \MCMCMC and \MCMC we compare the
stochastic error bars obtained after a given time (about one hour) on
a single CPU processor [see Fig.~\ref{fig_cdet_benchmark}]. For a
given order we run the \MCMCMC algorithm at order higher than the
standard MCMC as this is optimal for the \MCMCMC performance. Both
algorithms spend around $10\%$ of their time on normalisation. We
report an improvement of up to two orders of magnitude in the
computational time needed to reach a given stochastic error for the
highest expansion coefficient 12, and, importantly, we observe that
the improvement grows as a function of order.

\section{Conclusion}
In this \Article {$\;$}we have introduced a minimal generalization of the
Markov-Chain Monte Carlo algorithm that allows to consider multiple
configurations at the same Monte Carlo step. This new technique
permits straightforward parallelization, as well as giving the
possibility of using an arbitrary number of biased configurations at
the same Monte Carlo step.

After giving an intuitive picture of the algorithm based on a
graphical interpretation, we have presented numerical results for
spin-glass and quantum fermionic models, showcasing the generality and
the potential of this new technique. More specifically, we have shown
that, for the Sherrington-Kirkpatrick model, a very significant
reduction of thermalization time can be achieved when many
configurations are considered at each Monte Carlo step. We have
further shown how we can use the knowledge from all subsets of diagram
configurations within the Connected Determinant Diagrammatic Monte
Carlo method to further improve its computational efficiency, and we
have presented benchmark results for the paradigmatic Fermi-Hubbard model
away from half-filling where a speedup of up to two orders of
magnitude was observed at twelve Feynman loops.

We believe that our work has potential applications to areas where
Markov-chain Monte Carlo techniques are often considered too
computationally intensive due to their non-parallelizable nature, such
as big data Bayesian inference~\cite{mcmc_tall_data}. More generally,
the Markov-chain Monte Carlo algorithm for multiple configurations we
have introduced in this paper will be potentially applicable to
situations where the thermalization time is an issue and a parallel
computing machine is available, or when many biased configurations can
be evaluated at little computational effort.

\section*{Acknowledgments}
We thank Giuseppe Carleo, Michel Ferrero, Werner Krauth, Botao Li, and
Shiwei Zhang for valuable discussions.  We thank Dries Sels for
pointing out Ref.~\cite{coupling_mcmc}. This work has been supported
by the Simons Foundation within the Many Electron Collaboration
framework. The Flatiron Institute is a division of the Simons
Foundation. This work was granted access to the HPC resources of TGCC
and IDRIS under the allocations A0090510609 and A0070510609 attributed
by GENCI (Grand Equipement National de Calcul Intensif).

\appendix
\section{\label{sec_definition}
  General theory}

\subsection{Markov chain of configurations
\label{sub_mcmc_configuration}}
In this section  we review the standard MCMC algorithm. We consider a
discrete configuration space $\mathbb{S}$.  The Markov chain is
usually defined as a infinite sequence of configurations
$(c(j=1),\,c(j=2),...)$, $c(j)\in\mathbb{S}$, indexed by a discrete
``time'' $j$, such that a configuration $c(j+1)$ is generated
stochastically using only $c(j)$ as input.  Let us define $\Peq(c)$ as
the average time spent in the configuration $c\in\mathbb{S}$:
\begin{equation}\label{eq_def_peq_mcmc}
\Peq(c):=
\lim_{J\to\infty}
\frac{1}{J}
\sum_{j=1}^{J}
\delta_{c,c(j)}
\end{equation}
We also define $\Ptime{c}{j}$ as the probability of being in the
configuration $c$ at discrete time $j$.  We consider a Markov chain
such that (at least on average)
\begin{equation}
\lim_{j\to\infty }
\Ptime{c}{j} =
\Peq(c).
\end{equation}
The MCMC algorithm is particularly useful to compute average values of
an ``observable'' $\mathcal{O}$, which takes as arguments elements of the
configuration space $\mathbb{S}$:
\begin{equation}\label{eq_def_obs_conf}
\mathcal{O} :=
\frac{\sum_{c\in\mathbb{S}}\, w(c)
\, \mathcal{O}(c)}{\sum_{c\in\mathbb{S}}\, w(c)}
\end{equation}
where $w(c)\ge 0$ for every $c$.  If the condition
\begin{equation}\label{eq_p_as_mean}
\Peq(c)=
\frac{w(c)}
{\sum_{c'\in\mathbb{S}}w(c')}
\end{equation}
is satisfied for every $c\in\mathbb{S}$, then we can compute
$\mathcal{O}$ as a time average of $O(c(j))$, where $c(j)$ is the
Markov-chain state at discrete time $j$:
\begin{equation}\label{eq_monte_carlo_conf}
 \mathcal{O} =
  \lim_{J\to\infty}
  \frac{1}{J}
  \sum_{j=1}^{J}
  \mathcal{O}(c(j)).
\end{equation}

\subsection{Markov chain of graphs of configurations
\label{sub_mcmc_graph}}
In this section, we generalize the standard MCMC formalism of
Sec.~\ref{sub_mcmc_configuration} to to a Markov chain consisting of
an infinite sequence of undirected graphs of configurations
$(T(j=1),T(j=2),\dots)$, such that the set of nodes of $T(j)$,
$V(T(j))$, is contained in $\mathbb{S}$, and such that $T(j+1)$ is
stochastically generated using only $T(j)$ as input. We can introduce
the average time spent in a given graph $T\subseteq \mathbb{S}$
\begin{equation}\label{p_limit_set}
\Peq(T)=
\lim_{J\to\infty}
\frac{1}{J}
\sum_{j=1}^{J}
\delta_{T,T(j)}.
\end{equation}
At a continuous time $t\in[j, j+1[$, $j\in\mathbb{N}$, we imagine
visiting a configuration $c\in V(T(j))$ with a certain probability
$\Pcond{P}{(c,t)}{(T,j)}$ which we impose to be time-independent, and
we call it $\Pnode$:
\begin{equation}
\Pcond{\Pnode}{c}{T}
=:
\Pcond{P}{(c,t)}{(T,j)}.
\end{equation}
We assume this probability to be normalized to one: If the graph $T$
belongs to the Markov chain at time $j\in\mathbb{N}_0$, then for every $t\in[j, j+1[$, the system can be in only one node $c\in V(T)$ of the graph $T$
at at time, which is equivalent to imposing
\begin{equation}\label{eq_normalization_p_c_T}
\sum_{c\in V(T)} \,
\Pcond{\Pnode}{c}{T} = 1
\end{equation}
for every undirected graph $T$.

The Markov chain of undirected graphs is built with the sole purpose
of computing average values of functions defined in the nodes of the
graph. Therefore, if we define
\begin{equation}\label{eq_equilibrium_graph_conf}
\Peq(c) =
\sum_{T:\;V(T)\ni c}
\Peq(T)\;
\Pcond{\Pnode}{c}{T}
\end{equation}
and if we impose Eq.~\eqref{eq_p_as_mean}, we can compute observables
in configuration space from the Markov chain in graph space:
\begin{equation}\label{eq_monte_carlo_graph}
  \mathcal{O} =
    \lim_{J\to\infty}
    \frac{1}{J}
    \sum_{j=1}^{J}
    \sum_{c\in V(T(j))}
    \Pcond{\Pnode}{c}{T(j)}\;
    \mathcal{O}(c)
\end{equation}
where $\mathcal{O}$ is defined in Eq.~\eqref{eq_def_obs_conf}. The
advantage of this formulation compared to
Eq.~\eqref{eq_monte_carlo_conf} is that we can consider multiple
configurations at a given Monte Carlo step $j$, and visit each one of
them.

We now show how to compute $\Pnode$ from the graph transition
probability, which is a free parameter in this formulation. While we
could choose an arbitrary graph transition probability, here we limit
ourselves to the following choice: the new graph is generated starting
from one of the vertices of the old graph
\begin{equation}\label{eq_gen_graph}
\begin{split}
&\Pcond{P}{(T', j+1)}{(T,j)}:= \\
&\sum_{c\in V(T)\cap V(T')}
\Pcond{\Pnode}{c}{T}\,
\Pcond{\Pgen}{T'}{c},
\end{split}
\end{equation}
where $\Pcond{\Pgen}{T'}{c}$ is the probability of generating a graph
$T$ starting from a node $c$. $\Pgen$ can be chosen in a completely
arbitrary way, but a choice of $\Pgen$ constrains the form of
$\Pnode$. We will impose the detailed balance condition in
configuration space, which is defined as
\begin{equation}\label{eq_det_balance_graph}
P((c,t),\;(c',t')) =
P((c',t),\;(c,t')),
\end{equation}
where $c,c'\in\mathbb{S}$, $t<t'$, and $t,t'\in\mathbb{R}^+$ are much
larger than the thermalization time. First of all, we remark that in
our Monte Carlo process there are no transitions at non-integer times;
therefore, we can suppose $t'=j\in\mathbb{N}$, and $t=t'-0^+$. We then
obtain
\begin{equation}\label{eq_det_balance_graph2}
\begin{split}
&P((c,j-0^+),\;(c',j)) =
\sum_{T:\;V(T)\ni c}
\Peq(T) \;
\Pcond{\Pnode}{c}{T}\;\times \\
&
\sum_{T';\;V(T')\ni c'}
\Pcond{P}{(T',j)}{(T,j-1)}\;
\Pcond{\Pnode}{c'}{T}.
\end{split}
\end{equation}
 By using Eq.~\eqref{eq_gen_graph}, one can verify that the choice
\begin{equation}\label{eq_det_balance_graph_solution}
\Pcond{\Pnode}{c}{T}:=
\frac{\Peq(c)\;
\Pcond{\Pgen}{T}{c}
}{\Peq(T)}
\end{equation}
satisfies detailed balance [Eq.~\eqref{eq_det_balance_graph}]. We
remark that the probability of being in the undirected graph $T$,
$\Peq(T)$, does not need to be computed as one can use the
normalization condition [see Eq.~\eqref{eq_normalization_p_c_T}].

\subsection{Markov chain of sets of configurations
\label{sub_mcmc_set}}
In this section we consider a version of the Markov chain algorithm
for sets of configurations. We consider a Markov chain consisting of
an infinite sequence of sets of configurations
$(S(j=1),S(j=2),\dots)$, such that $S(j)$ is a set of fixed cardinality
$n\in\mathbb{N}$, $n>0$, contained in $\mathbb{S}$, and such that the
$S(j+1)$ is stochastically generated using only $S(j)$ as input.  We
can introduce the average time spent in a given set
$S\subseteq \mathbb{S}$ as
\begin{equation}\label{p_limit_set}
\Peq(S)=
\lim_{J\to\infty}
\frac{1}{J}
\sum_{j=1}^{J}
\delta_{S,S(j)}.
\end{equation}
At a continuous time $t\in[j, j+1[$, $j\in\mathbb{N}$, we visit a
subset $S'\subsetneq S$ with a certain probability
$P(\,(S',t)\,|\,(S,j)\,)$, which we impose to be time-independent:
\begin{equation}
\Pcond{\Psubset}{S'}{S}=:
\Pcond{P}{(S',t)}{(S,j)}.
\end{equation}
We assume this probability to be normalized to one: if $S$ is the
Markov-chain state at discrete time $j$, then we must be in one and
only subset $S'\subsetneq S$ for every continuous time $t\in[j,j+1[$,
which is equivalent to imposing
\begin{equation}\label{eq_normalization_p_S_Sp}
\sum_{S'\subsetneq S} \,
\Pcond{\Psubset}{S'}{S}=1
\end{equation}
for every $S\subseteq \mathbb{S}$ such that $|S|=n$. We would like to
evaluate
\begin{equation}\label{eq_observable_order}
\mathcal{O}_u :=
\sum_{S\subseteq \mathbb{S}:\;|S|=u}
w(S)\;
\mathcal{O}(S)
\end{equation}
for $u\in\{0,1,\dots, n\}$, and $w(S)\ge 0$. We introduce $\lambda_u>
0$ for $u\in\{0,\dots, n-1\}$, and define
\begin{equation}\label{eq_normalization_set}
\mathcal{N}_n:=
\sum_{S\subseteq \mathbb{S}:\; |S| < n}
\lambda_{|S|}\;w(S).
\end{equation}
We impose that the Markov chain thermalizes to the following
probability distribution for sets of cardinality less than $n$
\begin{equation}\label{eq_equilibrium_set}
\Peq(S)=
\frac{\lambda_{|S|}\;w(S)}{\mathcal{N}_n},
\end{equation}
for $S\subseteq \mathbb{S}$, $|S|< n$. We see that $\lambda_u$ is just
a reweighting parameter between different cardinalities $u$. We can
therefore estimate $\mathcal{O}_u$ as
\begin{equation}\label{eq_observable_monte_carlo_set}
\mathcal{O}_u=
\lim_{J\to\infty}
\frac{\mathcal{N}_n}{J}
\sum_{j=0}^{J-1}
\sum_{S'\subsetneq S(j), \;|S'|=u}
\frac{\Pcond{\Psubset}{S'}{S(j)}\;\mathcal{O}(S')}{\lambda_u},
\end{equation}
for $u\in\{0,1,\dots,n-1\}$, and
\begin{equation}\label{eq_observable_monte_carlo_set2}
\mathcal{O}_n=
\lim_{J\to\infty}
\frac{1}{J}\sum_{j=1}^{J}
\frac{w(S(j))\;\mathcal{O}(S(j))}
{|S|!\;\Peq(S(j))}.
\end{equation}
We see therefore that we can consider many subset configurations at
each Monte Carlo step.

We now show how to compute $\Psubset$ from the set transition
probability. We could choose an arbitrary set transition probability,
but we limit ourselves to the case where the new set is generated
starting from a subset:
\begin{equation}\label{eq_generation_set}
\begin{split}
&P((S',j+1)\,|\,(S, j)) =\\
&\sum_{S''\subseteq S\cap S'}
\Pcond{\Psubset}{S''}{S}\,
\Pcond{\Pgen}{S'}{S''}
\end{split}
\end{equation}
where $S\neq S'$, and $|S'|=|S|=n$.  The set generation function
$\Pgen$ can be chosen in an arbitrary way. As done in
Sec.~\ref{sub_mcmc_graph}, we impose detailed balance, that is, the
probability of being at $S$ at time $t$ and at $S'$ at time $t'$ is
equal to the probability of being in $S'$ at time $t$ and being in $S$
at time $t'$
\begin{equation}\label{eq_detailed_balance_set}
P((S, t),\;(S',t'))=
P((S', t),\;(S,t')),
\end{equation}
for $t<t'$, $t,t'\in\mathbb{R}^+$ are both much larger than the
thermalization time, and $S,S'\subseteq\mathbb{S}$, $|S|\le n$,
$|S'|\le n$. As there are no transitions at non-integer times, we can
just consider the case where $t=j-0^+$ and $t'=j$, where
$j\in\mathbb{N}$. Let us consider $S,S'\subseteq \mathbb{S}$ of
cardinality less than $n$, $|S|<n$, $|S'|<n$. One has
\begin{equation}\label{eq_detailed_balance_set2}
\begin{split}
&P((S,j-0^+),\;(S',j))=\\
&\sum_{S'':\;S\subsetneq S'',\;|S''|=n}
\Peq(S'')\;
\Pcond{\Psubset}{S}{S''}\;
\times \\
&\sum_{S''':\;S'\subsetneq S''',\;|S'''|=n}\;
P((S''',j)\,|\,(S'',j-1))\;\times\\
&\Pcond{\Psubset}{S'}{S'''}.
\end{split}
\end{equation}
By using Eq.~\eqref{eq_generation_set}, we can show that the
following choice for $\Psubset$ satifisfies detailed balance
[Equ.~\eqref{eq_detailed_balance_set}]:
\begin{equation}\label{eq_p_S_S}
\Pcond{\Psubset}{S}{S'}
=\frac{\Peq(S)\;
\Pcond{\Pgen}{S'}{S}}
{\Peq(S')},
\end{equation}
for $S,S'\subseteq \mathbb{S}$, $|S|<n$, $|S|'=n$. We can compute
$\Peq(S)$ for sets $S$ of cardinality $n$ by using
Equation~\eqref{eq_p_S_S} and the normalization condition
[Eq.~\eqref{eq_normalization_set}]
\begin{equation}\label{eq_P_eq_n}
\Peq(S)=\sum_{S'\subsetneq S}
\Peq(S')\;
\Pcond{\Pgen}{S}{S'},
\end{equation}
for $|S|=n$.

{\it Remark:} Using Eq.~\eqref{eq_p_S_S} and
Eq.~\eqref{eq_equilibrium_set}, we see that the explicit $\lambda_u$
factor in Eq.~\eqref{eq_observable_monte_carlo_set} simplifies to one
and we can safely take the $\lambda_u\to 0^+$ limit for some
$u\in\{0,1,\dots,n-1\}$. This means that we can accumulate statistics
for sets that are never visited by the Markov chain.

\section{\label{sec_generation_graphs}Generating graphs of configurations}
In this section we describe some choices for the graph-generation
probability $\Pgen$ introduced in Sec.~\ref{sub_mcmc_graph}. We
introduce two classes of graph-generation process: addition processes
[Sec.~\ref{sub_line_graph}], and populating processes
[Sec.~\ref{sub_gen_graph}].

\subsection{\label{sub_line_graph}
  Line-graph addition process}
 We consider here the following
process: starting from the configuration $c$, we create a new
configuration $c'$ with a probability given by some function
$\Pcond{\Pgen}{c'}{c}$, and we consider the graph with an edge between
$c$ and $c'$. Suppose now that we have created a graph consisting of
$c_0,c_1, \dots,c_{l-1}$ such that $c_k$ is connected by an edge to
$c_{k+1}$, for $k\in\{0,1,\dots,l-2\}$. With probability
$\frac{1}{2}$, we add a new vertex $c'$ connected to $c_0$ with the
probability distribution $\Pcond{\Pgen}{c'}{c_0}$, otherwise we add
the vertex $c'$ to the graph connecting it to $c_{l-1}$ with the
probability distribution $\Pcond{\Pgen}{c'}{c_{l-1}}$. If $T$ is the
line-graph with $c_0, c_1, \dots, c_{L-1}$ as nodes, one therefore has
\begin{equation}\label{eq_create_line_graph}
\begin{split}
\Pcond{\Pgen}{T}{c_k} =&
\frac{1}{2^L}
\left(\begin{array}{c}L  \\
k\end{array}\right)
\left(\prod_{k_l=0}^{k-1}
\Pcond{\Pgen}{c_{k_l}}{c_{k_l+1}}
\right)\;\times \\
&\left(\prod_{k_r=k+1}^{L-1}
\Pcond{\Pgen}{c_{k_r}}{c_{k_r-1}}
\right).
\end{split}
\end{equation}

\subsection{\label{sub_gen_graph}
  General graph populating process}
We now present a graph-generation process that is amenable to analytic
treatment.  The key point is that, if we first generate an ``empty''
graph, where no node is assigned, and we then populate the graph nodes
of with configuration values in such a way that the computation of
$\Pcond{\Pgen}{T}{c}$ becomes straightforward.  We consider
specifically a tree $T$.  We choose a root $r$ for the tree,
$r\in\{0,1,\dots,|V(T)|-1\}$, with a given probability distribution
$p_r$, which determines a rooted tree $R$.  From the root $r$, we can
populate the tree by generating a configuration for each edge of the
rooted tree with the probability distribution $\Pcond{\Pgen}{c}{c'}$,
where $c$ and $c'$ are connected with an edge going from $c$ to $c'$.
We can therefore write, assuming that $c$ is the root of the rooted
tree $R$
\begin{equation}\label{}
\Pcond{\Pgen}{T}{c} = 
p_r\,\prod_{(c,c')\in E(R)}
\Pcond{\Pgen}{c}{c'}
\end{equation}
One has complete freedom on $p_r$: the uniform choice is just
$p_r=\frac{1}{|V(T)|}$, which gives a self-thermalizing tree if
$\Pgen$ is symmetric.

\section{\label{sec_intro_diagmc}Introduction to Connected Determinant Diagrammatic Monte Carlo}
Below we provide a very brief overview of CDet.  We are interested in
computing a physical observable $\mathcal{C}(\xi)$, which we express
in terms of an infinite power series
\begin{equation}\label{cdet_sum}
  \mathcal{C}(\xi) := \sum_{n=0}^\infty \,c_n \, \xi^n,
\end{equation}
where the expansion coefficients $c_n$ are computed from the
stochastic sampling of multi-dimensional integrals over a set of $n$
internal variables $\{X_1,\dots,X_n\} =: S$ corresponding to vertex
positions in some arbitrary space:
\begin{equation}\label{cdet_integral}
  c_n =\frac{1}{N_n} \int_S\,\,c(S),
\end{equation}
where $N_n$ is a normalisation constant and $c(S)$ represents the sum
of weights for all topologically distinct connected graphs
$\mathcal{T}$ constructed from vertices in $S$ and obeying additional
rules imposed by the choice of model and observable. The weight of a
particular graph is given by the product of the weights of its edges
$\mathcal{E}(S)[i,j]$, which are functions of two vertex positions
$X_i$ and $X_j$. We can write:
\begin{equation}
  c(S)
  = \sum_{\mathcal{T}} \prod_{\{i,j\} \in \mathcal{T}} \mathcal{E}(S)[i,j].
\end{equation}
In general, the number of distinct connected graph topologies grows
factorially with the number of vertices and so does the computational
cost of evaluating $c(S)$. However, in many cases it is possible to
compute the related sum of the weights for all connected \emph{and}
disconnected graphs $a(S)$ at only polynomial or exponential
cost\footnote{This is often done by computing determinants or
permanents of $n\times n$ matrices $M(S)$ with edge weights as
entries: $\left(M(S)\right)_{ij} = \mathcal{E}(S)[i,j]$}.  In order to
obtain $c(S)$, one needs to eliminate all disconnected diagrams from
$a(S)$ using the recursive formula:
\begin{equation}\label{cdet_recursive_formula}
  c(S) = a(S) - \sum_{\substack{S^\prime \subsetneq S \\ S^\prime \ni
  X_1}} c(S^\prime) \, a(S \setminus S^\prime),
\end{equation}
where, in order to properly define connectivity, the sum is over all
subsets $S^\prime$ containing the arbitrarily chosen vertex $X_1$ from
$S$. The cost of evaluating Eq.\eqref{cdet_recursive_formula} scales
as $\mathcal{O}(3^n)$ with the number of vertices and, sometimes
together with the evaluation of $a(S)$, represents the computational
bottleneck of the CDet algorithm. It is important to note from
Eq.\eqref{cdet_recursive_formula} that in order to compute $c(S)$ for
a given set of vertices, one is obliged to compute $c(S')$ for all of
it's subsets $S'$, which can be used in computing lower order
expansion coefficients $c_{n'}$ where $n'<n$. This means that in the
process of computing the weight at order $n$ we are generating an
exponential ($2^n-1$) amount of a lower order weights as a side
product.

\section{\label{sec_generation_sets} Generating sets of configurations}

\subsection{\label{sub_create_set}
  Stochastic generation of the set}
In this section we define the stochastic process that we
perform to generate a set $S$ of cardinality $n \in \mathbb{N}_0$
given a subset $S'\subsetneq S$ and $|S'=m|$.
Let $S_0' = S'$.
Given $S_k'$, $k\in\{m,m+1,\dots, n\}$,
we generate $S_{k+1}'$ by adding a vertex to $S_k'$ in the following way:
We choose one element $e'$ of $S_k'$ randomly,
we generate a new element $e$ with the some arbitrary probability distribution $P_{\text{gen}}(e\;|\;e')$,
and we add it to $S_k'$ to generate $S_{k+1};=S_k\cup \{e'\}$.
We then identify $S_{n}'\equiv S$.

\subsection{\label{sub_rec_seed}
  Recursive exponential formula for the
generation probability}
We now show how to compute $P_{\text{gen}}(S|S')$ for $S'\subsetneq
S$, for a given $S$ of cardinality $n\in N_0$, $|S|=n$.  This is
needed to compute $P_{\text{subset}}$ [see Eq.~\eqref{eq_p_S_S}].  A
brute force approach would result in an algorithm that scales
factorially with $n$.  We are going to present a recursive formula
whose computational cost scales only exponentially with $n$.  We
define
\begin{equation}\label{eq_def_rec_P}
\tilde{P}_{S}(S'):=
P_{\text{gen}}(S\;|\;S\setminus S')
\end{equation}
for $S'\subsetneq S$.
One has
\begin{equation}\label{eq_rec_set_one_el}
\tilde{P}_{S}(\{e'\})=
\frac{1}{|S|}
\sum_{e''\in S\setminus \{e'\}}
P_{\text{gen}}(e'\;|\;e'')
\end{equation}
for $e\in S$.
For $S'\subsetneq S$, the Chapman-Kolmogorov equation can be written as
\begin{equation}\label{eq_rec_set_mul_el}
\begin{split}
\tilde{P}_{S}(S')=
\frac{1}{|S'|\,|S\setminus S'|}
\sum_{e''\in S\setminus S'}
\sum_{e'\in S'}&
P_{\text{gen}}(e'\;|\;e'')\\
&\cdot\;\tilde{P}_{S}(S'\setminus \{e'\})
\end{split}
\end{equation}
We can express the previous equation in words: {\it In order to
generate the subset $S$ from the set $S\setminus S'$, we need to first
generate an element $e'$ of $S'$ starting from one element $e''$ of
$S\setminus S'$, then generate the set $S$ from the set $S\setminus
S'\cup \{e'\}$}.  We note that that the cardinality of the l.h.s. of
Eq.~\eqref{eq_rec_set_mul_el} is one higher than the cardinality of
the r.h.s and therefore this equation can be solved recursively.  This
leads to a computational cost of $o(n^2\,2^n)$. It is possible to
improve upon this scaling by introducing a ``cumulative'' probability
distribution $\tilde{P}_{\text{gen}}(S',e')$ which we compute
recursively as:
\begin{equation}
\begin{split}
&\tilde{P}_{\text{gen}}(S',e') =\\
&\left\{ \begin{array}{ll}\sum_{e',e''\in S' e' \neq e''}
P_{\text{gen}}(e'\;|\;e'')
& \text{if}\;\;|S'|=1 \\ \tilde{P}_{\text{gen}}(S'\setminus \{e_{S'}\},e')
- P_{\text{gen}}(e_{S'}\;|\;e')
& \text{otherwise} \end{array} \right.  \end{split}
\end{equation}
where $e_{S'} \in S'$ can be chosen arbitrarily as long as
$e_{S'} \neq e'$. Then we can rewrite Eq.~\eqref{eq_rec_set_mul_el}
into
\begin{equation}\label{eq_rec_set_short}
\begin{split}
\tilde{P}_{S}(S')=
\frac{1}{|S'|\,|S\setminus S'|}
\sum_{e'\in S'}&
\tilde{P}_{\text{gen}}(S', e') \tilde{P}_{S}(S'\setminus \{e'\})
\end{split}
\end{equation}
which now has only computational cost of $O(n\,2^n)$. This
computational cost is negligible in comparison to $O(3^n)$, which is
the asymptotic computational scaling for the CDet~\cite{cdet}
algorithm used for the evaluation of connected Feynman diagrams.

\bibliographystyle{apsrev4-2}
\bibliography{main_biblio}

\end{document}